\documentclass{iopart}
\usepackage{graphicx} 
\usepackage{rotating}
\usepackage{cite}

\begin{document}

\title[Design of magnetic materials: Co$_2$Cr$_{1-x}$Fe$_{x}$Al.]
{Design of magnetic materials:\\
The electronic structure of the ordered, doped Heusler
compound Co$_2$Cr$_{1-x}$Fe$_{x}$Al }

\author{Gerhard H. Fecher$^1$, Hem Chandra Kandpal$^1$, Sabine Wurmehl$^1$,
        Jonder Morais$^2$, Hong-Ji Lin$^3$,
        Hans-Joachim Elmers$^4$, Gerd Sch\"onhense$^4$, and  Claudia Felser$^1$}
\address{$^1$ Institut f\"ur Anorganische Chemie und Analytische Chemie, \\
         Johannes Gutenberg - Universit\"at, D-55099 Mainz, Germany}
\address{$^2$ Instituto de Fisica, \\
         Universidade Federal do Rio Grande do Sul, Porto Allegre, 91501-970 RS, Brazil}
\address{$^3$ National Synchrotron Radiation Research Center - NSRRC, \\
         Hsinchu, 30076, Taiwan}
\address{$^4$ Institut f\"ur Physik, \\
         Johannes Gutenberg - Universit\"at, D-55099 Mainz, Germany}
\ead{felser@uni-mainz.de}

\date{\today}

\begin{abstract}
Doped Heusler compounds Co$_2$Cr$_{1-x}$Fe$_{x}$Al with varying Cr
to Fe ratio $x$ were investigated experimentally and theoretically.
The electronic structure of the ordered, doped Heusler compound
Co$_2$Cr$_{1-x}$Fe$_{x}$Al ($x=n/4, n=0,1,2,3,4)$ was calculated
using different types of band structure calculations. The ordered
compounds turned out to be ferromagnetic with small Al magnetic
moment being aligned anti-parallel to the $3d$ transition metal
moments. All compounds show a gap around the Fermi-energy in the
minority bands. The pure compounds exhibit an indirect minority gap,
whereas the ordered, doped compounds exhibit a direct gap. Magnetic
circular dichroism (MCD) in X-ray absorption spectra was measured at
the $L_{2,3}$ edges of Co, Fe, and Cr of the pure compounds and the
$x=0.4$ alloy in order to determine element specific magnetic
moments. Calculations and measurements show an increase of the
magnetic moments with increasing iron content. The experimentally
observed reduction of the magnetic moment of Cr can be explained by
Co-Cr site-disorder. The presence of the gap in the minority bands
of Co$_2$CrAl can be attributed to the occurrence of pure Co$_2$ and
mixed CrAl (001)-planes in the $L2_1$ structure. It is retained in
structures with different order of the CrAl planes but vanishes in
the $X$-structure with alternating CoCr and CoAl planes.
\end{abstract}

\pacs{75.50.Cc, 71.20.Lp, 78.70.Dm, 75.30.Cr}
\submitto{\JPCM}

\maketitle

\section{Introduction}

A great scientific interest is attracted by materials with a
complete spin polarization \cite{Pri98}. Such materials, being a
metal for spin up and a semiconductor (or insulator) for spin down
electrons, are called half-metallic ferromagnets \cite{GME83, CVB02}
(HMF). Heusler compounds have been considered potential candidates
to show this property \cite{GME83}. Theoretical calculations
predicted an energy gap for minority electrons for the half-Heusler
compound NiMnSb \cite{GME83, YM95} which, however, has been
controversially discussed \cite{RNB00a, RNB00b, ZSV01}. Similarly, a
HMF like behaviour was found by Plogmann \etal \cite{PSB99} for the
Cobalt based Heusler alloy Co$_2$MnSn.

Heusler compounds belong to a group of ternary intermetallics with
the stoichiometric composition X$_2$YZ ordered in an $L2_1$-type
structure, many of which are ferromagnetic \cite{WZ73}. X and Y are
transition metals and Z is usually a main group element. Y may also
be replaced by a rare earth element. The cubic structure consists of
four interpenetrating fcc lattices. The two fcc sub-lattices of the
X atoms combine to a simple cubic sub-lattice. Remarkably, the
prototype Cu$_2$MnAl is a ferromagnet even though none of its
constituents is one \cite{Heu03}. The $L2_1$ structure of the
Heusler compounds is shown in figure \ref{fig1}a.

\begin{figure}
\caption{Structure of ordered Co$_2$Cr$_{1-x}$Fe$_{x}$Al: \\
         a) $x=0, 1$ ($L2_1$), b) $x=1/4$ (3/4 if exchanging Fe and Cr),
         and c) $x=1/2$ (see text).}
\label{fig1}
\end{figure}

Ferromagnetic properties of various Heusler compounds have been
investigated experimentally \cite{WZ73, TFE74, EOS92, ZW74, EBE83,
YYG01} and theoretically \cite{IAK82, PSB99} (see Refs.\cite{LB19C,
LB32C} for comprehensive review). The Co-based Heusler compounds
Co$_2$YZ are of particular interest because they show a
comparatively high Curie temperature and varying magnetic moments
ranging from $0.3 \mu_B$ to $1.0 \mu_B$ at the Co site depending on
the constituents Y and Z (see Refs.: \cite{ZW74, Jez96, BNW00,
YIA02}).

The large variety of possible compositions of the Heusler compounds
allows easily to produce materials with predictable magnetic
properties. The easiest way to compose new materials is the exchange
of one or more of the elements X, Y, or Z. This is indeed widely
used in experiments and theory. See Refs.\cite{IAK82, IOK83, FSI90,
CSP96, IMF98, SJL98, PCF02, GDP02} for examples on Co$_2$ containing
Heusler compounds. However, the differences between the materials
are rather rough. A better fine-tuning of magnetic properties may be
obtained if substituting one or another constituent only partially.

Such a way to design new materials is possible using a deliberate
substitution of elements. Most oftenly the main-group element is
kept fixed because the magnetic properties are mainly governed by
the transition metal constituents. This leads to alloys of the type
X$_{2-x}$Y$_{1+x}$Z, (X$_{1-x}$X'$_x$)$_2$YZ or
X$_2$(Y$_{1-x}$Y'$_x$)Z ($x=0,..,1$). The first type is still a
ternary alloy whereas the second and third type result in quaternary
alloys. Co based alloys of the first two types have been
investigated by various groups \cite{EOS92, JB95, Jez96, SJN98,
STC99, LK02, Pug03}.

This work focuses on quaternary alloys of the third type. The
Cobalt-Aluminium based alloy of the X$_2$(Y$_{1-x}$Y'$_x$)Z type
with Y=Cr and Y'=Fe is particularly of interest as base for
materials design. This arises from the fact that the pure ($x=0$ or
1) compounds exhibit the same lattice parameter within 0.1\%.
Therefore, substituting partially one element by the other will lead
to a material with the same lattice parameter but changed electronic
and magnetic properties. Starting from the pure Cr containing
compound, the partial substitution by Fe may be seen as d-electron
doping.

Co$_2$Cr$_{0.6}$Fe$_{0.4}$Al is of special interest because a
relatively high magneto-resistance ratio of up to 30\% was found in
powder samples in a small magnetic field of $0.1 T$
\cite{BFJ03,FHK03}. Thin films of the compound were successfully
grown by several groups \cite{KC04,HKO05a,HKO05b,JCB05}. A
magneto-resistance ratio of 26.5\% \cite{IOG03} (at 5K) and 19\%
\cite{ITO04} (at room temperature) was found for a tunnelling
magneto-resistance (TMR) element of the same compound. Very
recently, Marukame \etal \cite{MKM05} reported a TMR ratio of 74\%
at 55K for a Co$_2$Cr$_{0.6}$Fe$_{0.4}$Al-MgO-CoFe magnetic tunnel
junction. A spin polarization of only less than 49\% was found for
poly-crystalline samples by means of Andreev reflections
\cite{AJB03}. The observation of an incomplete spin polarization may
not only be caused by the model used to interpret the data
\cite{AJB03,CFJ05} but also by the properties of the sample.
Clifford \etal \cite{CVG04} reported recently a spin polarisation of
81\% in point contacts of Co$_2$Cr$_{0.6}$Fe$_{0.4}$Al.


For the purpose of the present study, doped Heusler alloys
Co$_2$Cr$_{1-x}$Fe$_{x}$Al were prepared by arc-melting under an
argon atmosphere. The resulting specimens were dense
poly-crystalline ingots. Structural properties were measured using
X-ray diffraction as a standard method. The cubic structure with a
lattice constant of about $5.73$\AA was confirmed for all samples.
Flat discs ($8 mm$ diameter by $1 mm$ thickness) were cut from the
ingots. The discs were mechanically polished for spectroscopic
experiments.

Field dependent magnetic properties were measured by SQUID -
magnetometry (temperature: $4K$ to $300 K$) and by the
magneto-optical Kerr effect (MOKE) at room temperature. The remanent
magnetization was less than $10\%$ of the saturation magnetization
pointing on a soft magnetic material. Saturation was achieved for
external fields above $0.2 T$ at $300 K$. Especially the Co$_2$CrAl
samples showed large differences in the total moment varying from
$1\mu_B$ to $3\mu_B$ per formula unit. This depends mostly on the
post-processing of the samples like annealing followed either by
quenching or slow cooling at different rates. In some cases, X-ray
diffraction exhibited pronounced super-structures pointing on a
tetragonal distortion of the unit cell. It should be noted that some
types of disorder cannot be detected easily by X-ray powder
diffraction as the scattering coefficients of Co and Cr are very
similar. The same applies for neutron diffraction. Due to the nearly
equal scattering length of Cr and Al, in particular, it is not
possible to distinguish between ordered $L2_1$ and disordered $B_2$
structures. Therefore, we will use a detailed analysis of the
magnetic properties to gain information about structural disorder of
the samples. It should be noted that a mixing of Cr and Fe atoms in
doped compounds will be hardly detectable by X-ray diffraction. This
is caused by the very similar scattering coefficients of the
constituting elements. X-ray magnetic circular dichroism (MCD) in
soft X-ray absorption spectroscopy was performed at the {\it First
Dragon} beamline of NSRRC (Hsinchu, Taiwan). The MCD measurements at
the Cr, Fe and Co $L_{2,3}$ absorption edges for
Co$_2$Cr$_{1-x}$Fe$_{x}$Al ($x=0$, $0.4$, and $1$) were carried out
in order to investigate element specific magnetic properties and
compare them with theoretical predictions. A Co$_2$CrAl sample free
of super-structure but with too low magnetic moment was selected for
the MCD measurements in order to explain the large deviation from
the expected value of $3\mu_B$ per formula unit.
Co$_2$Cr$_{0.6}$Fe$_{0.4}$Al was selected for the MCD measurements
as this composition has shown the largest effect in measurements of
the magneto resistance.

More details about the experiment and the data analysis are reported
in Ref.\cite{EFV03,FHK03}.

The present work reports on calculations of the electronic and
magnetic properties of ordered Heusler compounds of the type
X$_2$(Y$_{(1-i/4)}$Y'$_{i/4}$)Z. The calculated properties are
compared to experimental values. Deviations from the $L2_1$
structure are discussed on hand of ordered structures. Random alloys
of the X$_2$(Y$_{(1-x)}$Y'$_x$)Z type with non-rational values of
$x$ as well as random disorder (for examples see references
\cite{MNS04,KUK04}) will not be discussed here.

\section{Calculational Details}

Self-consistent band structure calculations were carried out using
the scalar - relativistic full potential linearised augmented plane
wave method (FLAPW) provided by Blaha \etal \cite{BSS90,BSM01} (Wien2k).
The exchange-correlation functional was taken within the generalized
gradient approximation (GGA) in the parametrization of Perdew \etal
\cite{PCV92}. For comparison, calculations were also performed using
the linear muffin-tin orbital (LMTO) method provided by Savrasov
\cite{Sav96} (LMTART 6.5) on different levels of sophistication from
simple atomic sphere approximation (LMTO-ASA) to full potential
plane wave representation (FP-LMTO-PLW). A $20\times20\times20$
k-point mesh was used for the integration in cubic systems.

The properties of the pure Cr or Fe containing compounds where
calculated in $F\:m\overline{3}m$ symmetry using the experimental
lattice parameter ($a=10.822 a_{0B}$, $a_{0B} = 0.529177$\AA ) as
determined by X-ray powder diffraction. All muffin tin radii where
set to nearly touching spheres with $r_{MT}=2.343 a_{0B}$ in both
full potential methods. The overlapping spheres where set to
$r_{MT}=2.664 a_{0B}$ for the ASA calculations.

The full formula sum of the cubic cell is X$_8$Y$_4$Z$_4$, with X =
Co, Y = Cr or Fe and Z=Al and is reduced to X$_2$YZ = Co$_2$YAl.
Exchanging Y and Z leads indeed to identical structures. (See figure
\ref{fig1} and table \ref{Tab1} for the positions of the atoms.)

The calculation of mixed random alloys is not straight forward in
both (FLAPW and LMTO) calculational methods. However, substituting
some Cr atoms of the $L2_1$ structure by Fe leads in certain cases
to ordered structures that can be easily used for calculations.
Ordered, mixed compounds may have the general formula sum
X$_8$(Y$_{(1-x)}$Y'$_x$)$_4$Z$_4$ with Y = Cr and Y' = Fe. These
structures have integer occupation of Y and Y' if $x=i/4$ with
$i=1,2,3$.

Start with the Cr atoms occupying the corners of the cube and the
centre of the faces. The Al atoms are located at the middle of the
cube-edges (see: figure \ref{fig1}a). Replacing the Cr atom at
(0,0,0) by Fe leads to the structure with $x=1/4$ (see: figure
\ref{fig1}b). The same structure may also be found if starting with
Cr and Al exchanged and then replacing the Cr atom at (1/2,1/2,1/2)
by Fe. The symmetry of these structures is again cubic but reduced
to $P\:m\overline{3}m$. The only difference is that the base atoms
are shifted.

Exchanging simply the Cr atoms of this structure by Fe leads to the
structure with $x=3/4$. The accompanied formula sums are
Co$_8$Cr$_3$FeAl$_4$ and Co$_8$CrFe$_3$Al$_4$.

Again, start with the Cr atoms occupying the corners and
face-centres of the cube. Replace two of the face Cr atom, say at
(0,1/2,1/2), (1/2,0,1/2) and at opposite faces by Fe. The result is
the structure with $x=1/2$. This structure is initially cubic but
can be reduced to tetragonal symmetry ($P\:4/mmm$) with the formula
sum Co$_4$CrFeAl$_2$.

The three different structures are illustrated in figure \ref{fig1}.
The cell shown for $x=1/2$ has a reduced, tetragonal symmetry. The
z-axis may be chosen such that the long c-axis coincides with one of
the cubic axes of the initial structure. The symmetry and the
lattice sites of the structures are summarized in table \ref{Tab1}.

\begin{table}
    \caption{Symmetry. \\
             Space groups, and Wyckoff positions
             of the constituents in ordered Co$_2$Cr$_{1-x}$Fe$_x$Al.}
    \begin{center}
        \begin{tabular}{llcccc}
        compound             &                     & Co & Cr & Fe & Al \\
        \hline
        Co$_2$CrAl           & $F\:m\overline{3}m$ & 8c & 4a & -  & 4b \\
        Co$_8$Cr$_3$FeAl$_4$ & $P\:m\overline{3}m$ & 8g & 1a & 3c & 1b, 3d \\
        Co$_4$CrFeAl$_2$     & $P\:4/mmm$          & 4i & 1a & 1d & 1b, 1c \\
        Co$_8$CrFe$_3$Al$_4$ & $P\:m\overline{3}m$ & 8g & 3c & 1a & 1b, 3d \\
        Co$_2$FeAl           & $F\:m\overline{3}m$ & 8c & -  & 4a & 4b
        \end{tabular}
    \end{center}
    \label{Tab1}
\end{table}

Other ordered structures are found from larger elementary cells. The
cubic cell doubled in all three directions has the overall formula
sum X$_{64}$(Y$_{(1-x)}$Y'$_x$)$_{32}$Z$_{32}$. The special case
with $x=13/32$ is very close to the compound
Co$_2$Cr$_{0.6}$Fe$_{0.4}$Al.

\section{Results and Discussion}

The electronic structure of the pure and doped compounds will be
discussed in the following. First, the band structure and the
density of states of the ordered compounds are presented. This is
followed by a more specific discussion of the magnetic properties on
hand of measured and calculated magnetic moments.

A structural optimization was performed for Co$_2$CrAl and
Co$_2$FeAl using FLAPW in order to verify using experimental lattice
parameter. The energy minima were found to appear at lattice
parameter being less than 0.5\% smaller compared to the experimental
values, in both materials. The calculated bulk moduli were 217~GPa
and 210~GPa for the Cr and the Fe containing compounds,
respectively. None of the results discussed below changes
significantly if using the optimized lattice parameter instead of
the experimental one. In particular, the overall spin moments stay
the same and the half-metallic behavior of the ordered compounds
retains. Very small deviations appear for elemental resolved values.
Those are already sensitive to the setting of the $r_{MT}$ and the
number of k-points used for integration, as is well known. The only
small differences in the observed and optimized structures do not
allow to notice any deviation from Vegards law for the mixed
compounds.

\subsection{Band structure and density of states}

\begin{figure}
\includegraphics[width=12cm]{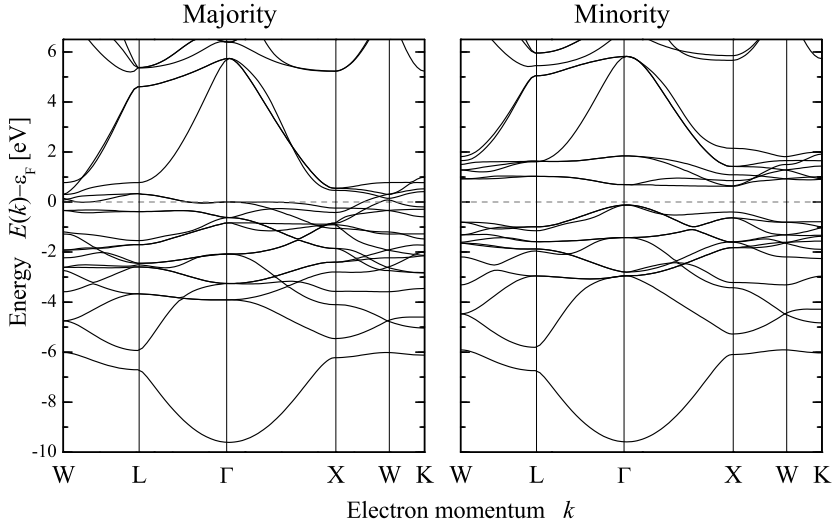}
\caption{Band structure of Co$_2$CrAl.}
\label{fig2}
\end{figure}

The self-consistent FLAPW band structure of Co$_2$CrAl is shown in
figure \ref{fig2}. The energy scale is referenced to the
Fermi-energy ($\epsilon_F$). The typical Heusler gap is located at
about $6eV$ binding energy. It separates the low lying s bands from
bands of predominately d character. These low lying s bands emerge
mainly from the main group element, here Al. This gap is very small
in the Al containing compounds. Much larger gaps are found for
example in Sn containing compounds like Co$_2$TiSn \cite{TPK96} or
the half-Heusler NiMnSb \cite{GME84}.

From the spin resolved bands, it is seen that the majority bands
cross or touch the Fermi-energy ($\epsilon_F$) in rather all
directions of high symmetry. On the other hand, the minority bands
exhibit a gap around $\epsilon_F$ thus confirming a HMF character.
For Co$_2$CrAl, the width of the gap is given by the energies of the
highest occupied band at the $\Gamma$-point and the lowest
unoccupied band at the $\Gamma$ or $X$-point. The smaller value is
found between $\Gamma$ and $X$, thus it is an indirect gap. It
should be noted that the direct gap at the $\Gamma$-point is only
$60meV$ wider. Therefore, a small change in the parameters of the
calculation may already change the character of the gap from
indirect to direct. Indeed, some LMTO calculations resulted in a
direct gap, most probably just for that reason.

We will restrict the following comparison of the doped compounds to
the $\Delta$-direction being parallel to [001]. The
$\Delta$-direction possesses in all cases $C_{4v}$ symmetry. It has
the advantage that the compound with $x=1/2$ can be compared
directly to the others even so it is calculated for tetragonal
symmetry where the corresponding $\Lambda$-direction is between
$\Gamma$ and $Z$.

The $\Delta$-direction is perpendicular to the Co$_2$ (100)-planes.
As will turn out later, just the $\Delta$-direction plays the
important role for the understanding of the HMF character and
magnetic properties of Heusler compounds. This role of the
$\Delta$-direction was also pointed out by \"O\^g\"ut and Rabe
\cite{OR95}.

\begin{figure}
\includegraphics[width=7cm]{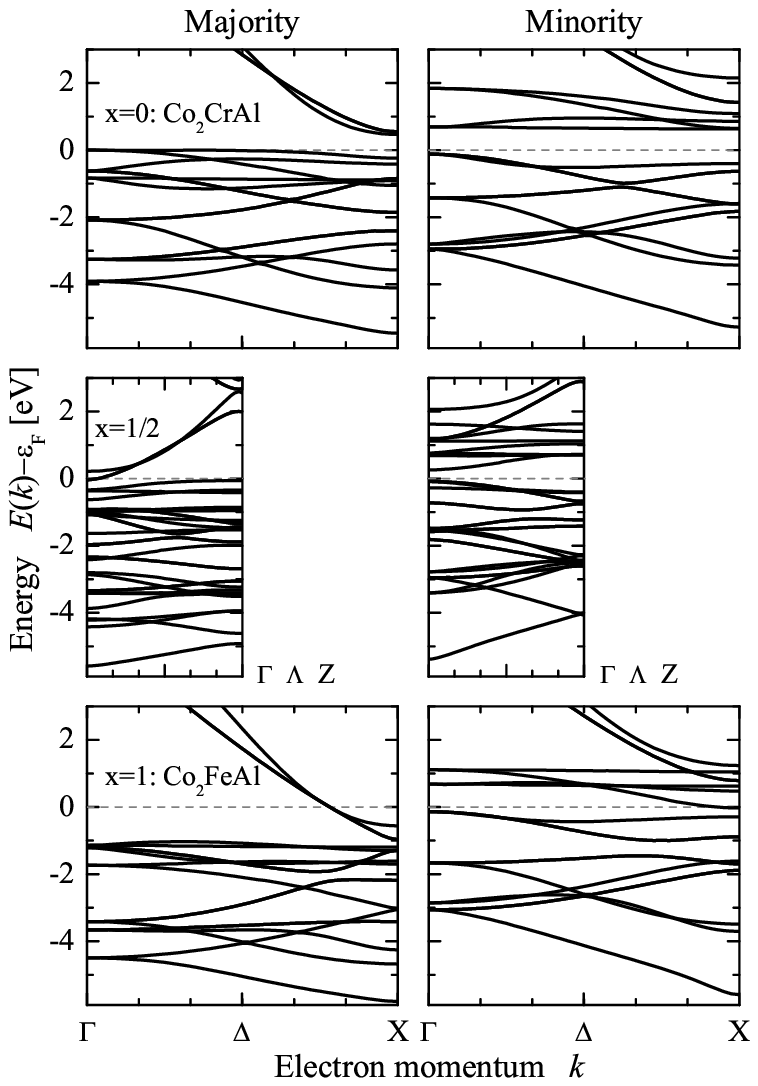}
\caption{$\Delta$,$\Lambda$-bands of Co$_2$Cr$_{1-x}$Fe$_x$Al. \\
(The high symmetry points are $X$ for $x=0, 1$ and $Z$ for
$x=1/2$. Note that the absolute values of $k$ at $X$ are:
1.1\AA$^{-1}$ for $x=0, 1$ and only 0.55\AA$^{-1}$ at $Z$ for
$x=1/2$.)}
\label{fig3}
\end{figure}

The band structures in $\Delta$-direction of the pure ($x=0, 1$) and
the doped ($x=1/2$) compounds are displayed in figure \ref{fig3} for
energies above the Heusler gap. In general, the doped compounds
exhibit much more bands compared to the pure ones as a result of the
lowered symmetry. Therefore, results are shown only for the mixed
compound with equal Fe and Cr concentration.

Compare the majority bands of the two pure compounds. At a first
sight, the Fermi-energy is just higher in the Fe case compared to
Cr, as expected from the larger number of d-electrons. A closer look
reveals more detailed differences. The indirect gap of the
$\Delta$-direction (clearly seen for the Cr based compound) is not
only shifted below $\epsilon_F$ but nearly closes for the Fe based
compound. This gap is also nearly closed in the majority bands of
the mixed compound with $x=1/2$ as well as those with $x=1/4, 3/4$
(not shown here). This observation calls a rigid band model in
question. In that case the bands would be simply filled with
increasing number of d-electrons, leaving the shape of the bands
unchanged.

More interesting is the behaviour of the minority bands as those
determine the HMF character of the compounds. Comparing again the Cr
and the Fe based compounds, one finds that the energies of the
states at $\Gamma$ are nearly the same below $\epsilon_F$. The
shapes of the bands close to $\Gamma$ are similar, too. The
situation is different at $X$ where the unoccupied states are
shifted toward $\epsilon_F$ in Co$_2$FeAl compared to Co$_2$CrAl.

It is worthwhile to note that the first unoccupied minority band of
Co$_2$FeAl just touches $\epsilon_F$ at $X$. Therefore, any
temperature above $0K$ will immediately destroy the HMF gap due to
the smearing of the DOS around $\epsilon_F$ (for additional
temperature effects destroying the minority gap see
e.g.:\cite{SD02,DS03}). Chioncel \etal \cite{CKG03} reported for
NiMnSb the occurrence of nonquasiparticle states just below the
minority conduction band. A similar effect would immediately destroy
the HMF character if appearing in Co$_2$FeAl, too. The band
structure of Co$_2$MnAl shows a similar behaviour at $\Gamma$ like
Co$_2$FeAl. Here the minority bands are even crossing slightly the
Fermi-energy, as was found in calculations for comparison with the
iso-electronic compound Co$_2$Cr$_{1/2}$Fe$_{1/2}$Al. Even if
accounting for small numerical deviations while calculating
$\epsilon_F$ for the two compounds, they may not be good candidates
for spin-injection devices.

The mixed compounds are described by $P$ lattices. Therefore, the
Brillouin zone of these compounds is generally smaller compared to
the $F$ lattice. This results in a seemingly back-folding of bands
from the larger $F$-Brillouin zone into a smaller one. This effect
is accompanied with some additional splitting (removed degeneracies)
at points of high symmetry.

Due to the manifold of bands in the mixed compounds, it is not easy
to compare the results directly, therefore we concentrate on the
width of the gap in the minority bands. This gap is mainly
characterized by the bands in $\Delta$-direction as found from the
band structure for all directions of high symmetry (not shown here).
More specifically, it is given by the energies at the $\Gamma$ and
$X$ points of the Brillouin zone.

\begin{figure}
\includegraphics[width=7cm]{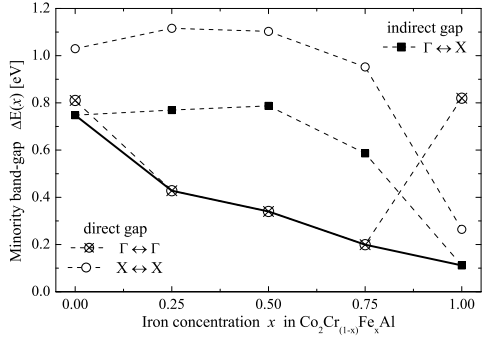}
\caption{Minority Band Gap in Co$_2$Cr$_{1-x}$Fe$_{x}$Al.\\
        (Lines are drawn to guide the eye. The full drawn line follows the
         limit for the HMF-gap)}
\label{fig4}
\end{figure}

The width of the gap in the minority bands is shown in figure
\ref{fig4}. The direct band gap at the $\Gamma$-point becomes
successively smaller with increasing iron content $x$ and ranges
from $750meV$ at $x=0$ to $110meV$ at $x=1$. The direct gap of
Co$_2$FeAl is only $60meV$ wider than the indirect one between
$\Gamma$ and $X$. The direct gap at $\Gamma$ is much wider in
Co$_2$FeAl, therefore this compound is characterized by the indirect
gap only.

The character of the gap changes from indirect to direct if
comparing pure and mixed compounds, respectively. This change of the
character of the gap in the $\Delta$-direction is a consequence of
the smaller Brillouin zone in the mixed compounds that leads to a so
called back-folding of bands.

\begin{figure}
\includegraphics[width=7cm]{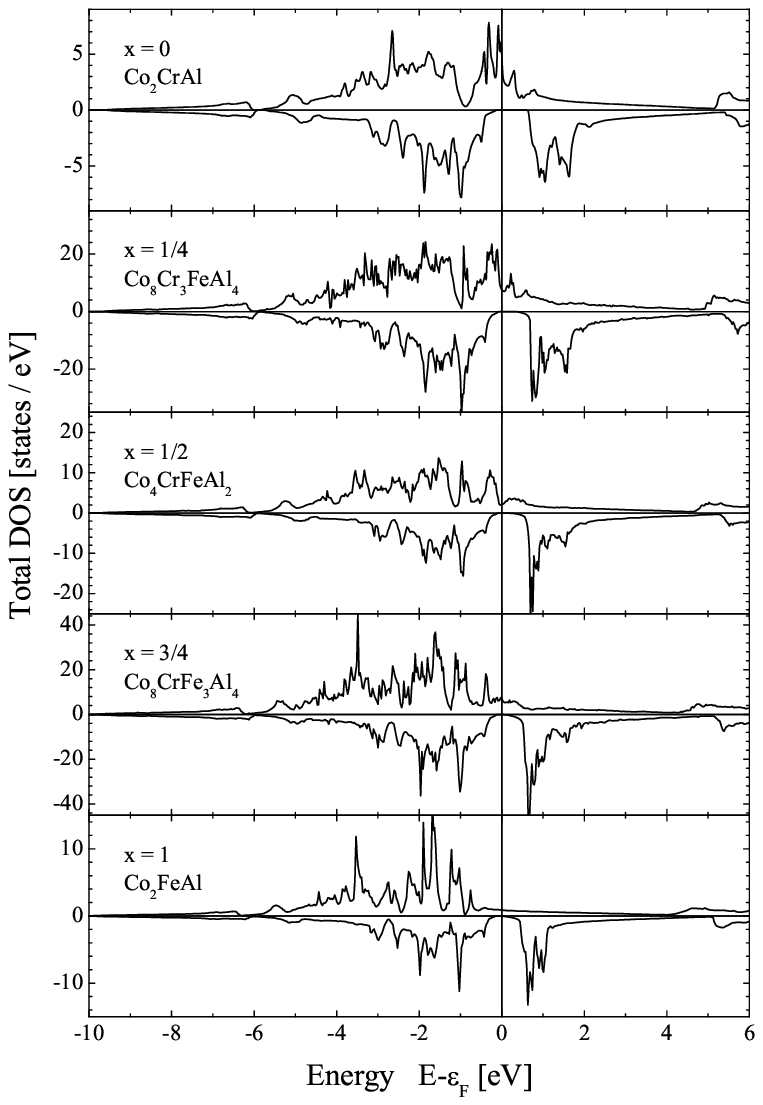}
\caption{Total DOS of Co$_2$Cr$_{1-x}$Fe$_{x}$Al.}
\label{fig5}
\end{figure}

The total density of states (DOS) is shown in figure \ref{fig5} for
varying iron content $x$. The gap at the Fermi-energy is clearly
seen in the DOS of the minority states for all compounds. The total
DOS shows also that the Heusler-gap at about $6eV$ binding energy is
nearly closed.

The majority DOS at the Fermi energy decreases with increasing iron
concentration $x$. The density of majority electrons at $\epsilon_F$
is a crucial point for spectroscopic methods investigating the spin
polarization, like spin-resolved photoemission. A complete spin
polarization may be only detectable if there is a high majority
density. The same may be true for spin injection systems where one
is interested in a high efficiency.

It is also seen that the minority DOS seems to be much less effected
by the Fe doping compared to the majority DOS. Mainly the unoccupied
part of the DOS above $\epsilon_F$ changes its shape but not the
occupied part.

In summary, it is found that doping of the compound with Fe changes
mainly the occupied majority and the unoccupied minority DOS. Again
it is seen that doping by iron results not just simply in a shift of
the DOS as would be expected from a rigid band model. Majority and
minority densities are altered in a different way.

\begin{figure}
\includegraphics[width=12cm]{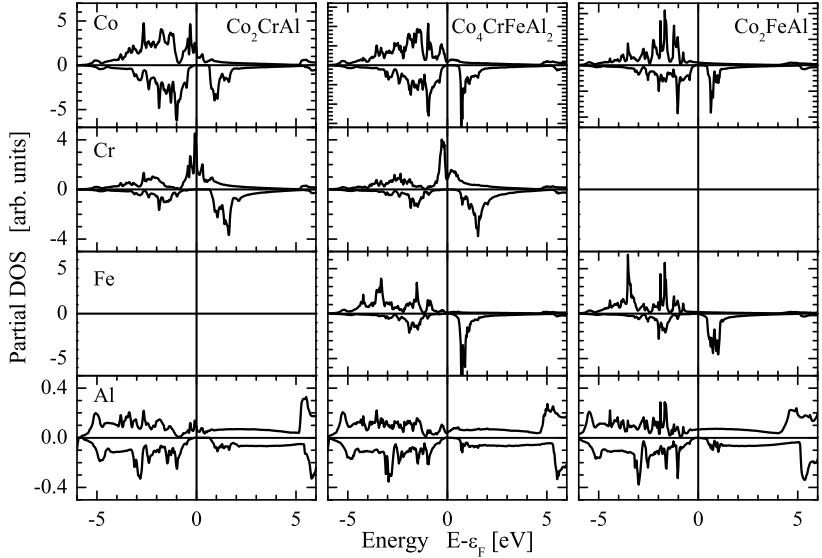}
\caption{Partial DOS of Co$_2$Cr$_{1-x}$Fe$_{x}$Al ($x=0, 1/2, 1$).}
\label{fig6}
\end{figure}

More details of the change in the DOS and electronic structure can
be extracted if analysing the partial DOS (PDOS), that is the atom
type resolved density of states. The PDOS of the pure compounds is
compared in figure \ref{fig6} to the PDOS of the mixed compound with
equal Cr and Fe content ($x=1/2$).

From figure \ref{fig6}, it is seen that the high majority DOS at the
Fermi-energy emerges from Cr. Both, Co and Fe exhibit only a small
majority PDOS at $\epsilon_F$. Overall, the change of the majority
DOS of Co$_2$Cr$_{1-x}$Fe$_{x}$Al around $\epsilon_F$ can be clearly
attributed to the increasing amount of iron with respect to
chromium. The minimum in the minority DOS around $\epsilon_F$ is
mainly restricted by the shape of the Co PDOS. This indicates that
the HMF like behaviour is mainly characterized by Co. The steep
increase of the minority PDOS of Cr and Fe is mainly located in the
unoccupied part above $\epsilon_F$.

Doping with Fe does not only change the total DOS but also the PDOS
of Co and Cr. In particular, the slight shift of the Cr PDOS to
lower energies causes an additional decrease of majority states at
$\epsilon_F$. This shift increases with increasing Fe concentration
as was found from the PDOS for $x=1/4$ and $3/4$ (not shown here).
The slight energy shift of the PDOS will result in a small change of
the local magnetic moments at the Co and Cr sites, as will be shown
below.

The aluminium PDOS stays rather unaffected from the Cr or Fe
concentration. It exhibits only small energy shifts.

\subsection{Magnetic moments}

\begin{figure}
\includegraphics[width=7cm]{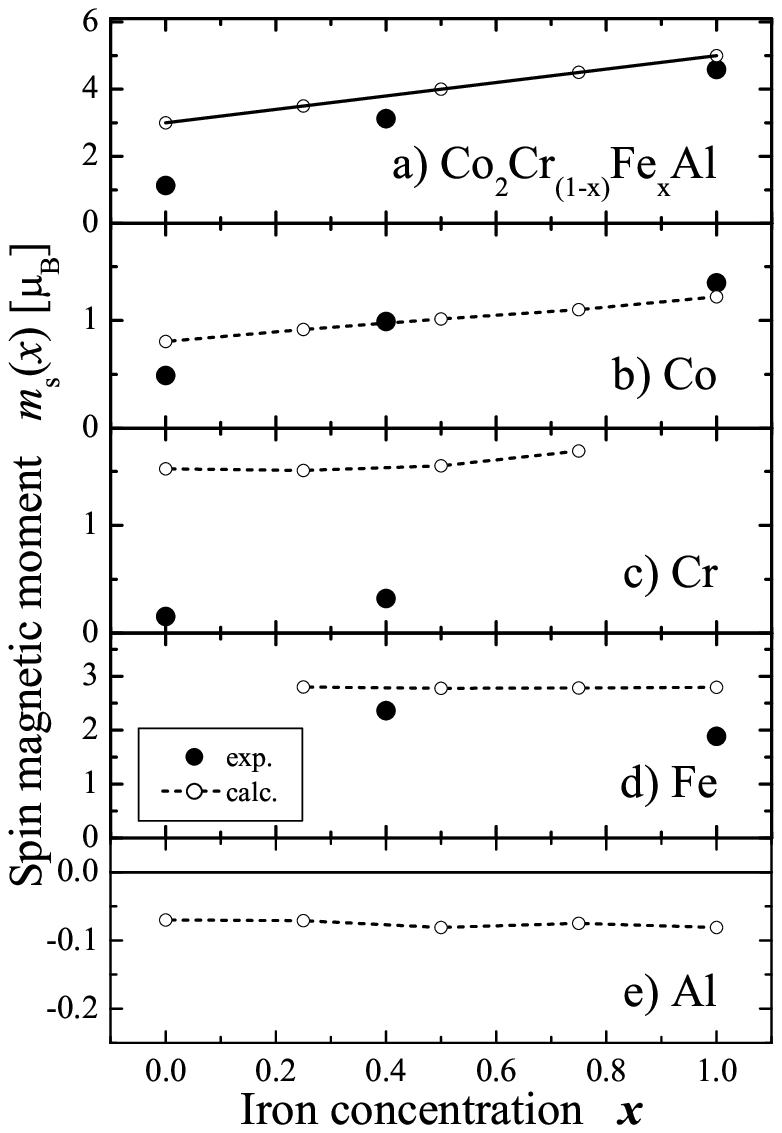}
\caption{Magnetic moments of Co$_2$Cr$_{1-x}$Fe$_{x}$Al.\\
         The measured element specific and total moments for ($x=0$, $0.4$,
         and $1$) are compared to calculated values for ($x=0$, $1/4$,
         $1/2$, $3/4$, and $1$). The full line in a) corresponds to the
         {\it thumb-rule}, dashed lines in b)...e) are drawn through the
         calculated values to guide the eye.}
\label{fig7}
\end{figure}

SQUID magnetometry and MCD were used to determine total and partial
magnetic moments of Co$_2$Cr$_{1-x}$Fe$_{x}$Al. Details of these
measurements are reported in Ref.\cite{EFV03}. The measured values
are extrapolated to $T = 4K$ and calibrated using the total moments
measured by SQUID. Total and element specific magnetic moments were
extracted from the band structure calculations reported above.
Figure \ref{fig7} compares measured and calculated spin magnetic
moments.

The calculated total spin magnetic moment follows the {\it rule of
thumb} for Heusler compounds:

\begin{equation}
    \mu_s = (N-24)\mu_B
\label{eq1}
\end{equation}

$N$ is the cumulated number of valence electrons (here: 4s, 3d for
the transition metals Co, Cr, or Fe and 3s, 3p for the main group
element Al). The value calculated from equation \ref{eq1} is shown
as full line in figure \ref{fig7}a.

The calculated spin magnetic moments of Co and Fe are in agreement
with the measured values. The measured, lower total value at small
Fe concentration can be attributed clearly to a too low moment at
the Cr sites.

The calculated spin moments of Co and Cr increase slightly with
increasing Fe concentration, whereas the Fe moment stays nearly
constant. This increase is explained by an energy shift of the
partial densities of Co and Cr as discussed above.

The calculated Al spin moment is negative, independent of the Fe
concentration. This points on an {\it anti-ferromagnetic} order of
the Al moments with respect to the transition metal moments.
However, those induced moments at the Al sites are only very small.

All values found here for the pure compounds are in the same order
of those calculated by Galanakis et.al.\cite{GDP02} using the
Korringa-Kohn-Rostokker (KKR) method.

It may be interesting to compare the magnetic moments of
Co$_2$Cr$_{0.5}$Fe$_{0.5}$Al with those of the nominally
iso-electronic compound Co$_2$MnAl. Our calculations yielded
$3.81\mu_B$ for the total spin moment per formula unit and
$0.84\mu_B$ and $2.58\mu_B$ for the partial spin moments of Co and
Mn, respectively. These values are similar to those found by other
groups\cite{GDP02, FSI90} using KKR. The Co moment is larger in the
mixed compound ($1.01\mu_B$) and the average Cr$_{0.5}$Fe$_{0.5}$
moment ($2.24\mu_B$) is smaller compared to Mn. It is interesting to
note that the minority states were shifted at the $\Gamma$-point to
energies above $\epsilon_F$, such that the HMF-gap became closed in
Co$_2$MnAl. A variation of the lattice parameters showed that this
is not causing the differences between the two materials. The
differences are caused by the different local potentials in those
compounds.

Most evidently, the calculated magnetic moment of the Co$_2$CrAl
compound does not agree with the measured one. The experimental
value is only about $1.3\mu_B$ whereas theory predicts a value of
about $3\mu_B$ per formula unit. The value found here is in rough
agreement to the ground state magnetic moment of $1.55\mu_B$ per
formula unit reported earlier by Buschow\cite{BE81}. It was
previously considered that mainly the Co atoms carry the magnetic
moment, whereas the contribution of Cr and Al atoms remains
negligible \cite{SNS01}. This empiric assumption may describe the
present element specific measurements revealing a moment of $\approx
2\times0.55\mu_B$ for Co$_2$ but only $\approx 0.2\mu_B$ for Cr.
However, it does not explain the physics behind that observation.

One may assume that the high magnetic moment of Cr is an artefact of
a particular calculation scheme. Therefore, different calculation
schemes were used to check a probable occurrence of such effects.
The results are summarized in table \ref{Tab2}. The results of
Galanakis \etal \cite{GDP02} derived from KKR-calculations are shown
for comparison. The partial and total magnetic moments calculated by
Kellou \etal \cite{KFG03} at a smaller lattice parameter
(10.758$a_{0B}$) are in the same range.

\begin{table}
    \caption{Magnetic moments in Co$_2$CrAl.\\
    Element specific (per atom) and total spin magnetic moments
    (per formula unit) calculated by different methods.}
    \begin{center}
        \begin{tabular}{lccc}
        \\
                        & & $m_{s} [\mu_{B}]$ & \\
        Method          & Co   & Cr   & tot \\
        \hline
        LMTO-ASA        & 0.67 & 1.72 & 2.98 \\
        LMTO-ASA-GGA    & 0.70 & 1.63 & 2.98 \\
        FP-LMTO-PLW     & 0.73 & 1.61 & 3.01 \\
        FLAPW-GGA       & 0.81 & 1.52 & 3.00 \\
        KKR\cite{GDP02} & 0.76 & 1.54 & 2.96
        \end{tabular}
    \end{center}
    \label{Tab2}
\end{table}

It is seen from table \ref{Tab2} that all values stay comparable
within few \%. Therefore, the deviation from the experiment cannot
be attributed to the peculiarities of one or another theoretical
method. The Cr moment is about $(1.6\pm0.1)\mu_B$ rather independent
of the method of calculation.

On the other hand, a reduction of the observed Cr moment may be
caused by site-disorder, such that part of the Cr atoms become
anti-ferromagnetically ordered either among each other or with
respect to the Co atoms. The latter results in a ferrimagnetic
order. A mixture of ordered and disordered crystallites will then
result in a measured value being too small compared to the
calculated moment. Such a disorder has to be realized with the same
lattice parameter because X-ray diffraction did not show any
superstructure in the samples investigated here.

To estimate the influence of disorder in the CrAl-planes, the band
structure was also calculated for the hypothetical $cP16$ ferrite
structure ($P\:m\overline{3}m$). In this structure, successive
CrAl-(001)-planes are Cr or Al rich with $3:1$ or $1:3$ ratios,
respectively. The calculated magnetic moments of Cr were about
$1.5\mu_B$. Lin and Freeman \cite{LF92} used the tetragonal $tP4$
structure ($P\:4/mmm$) to simulate a Y/Al disorder in Ni$_2$YAl
(Y=Ti, V, etc.). In this structure one has the series Co-Cr-Co-Al
for the consecutive (001)-planes instead of Co-(CrAl) (Note that
(100) and (001) planes are not equivalent in the tetragonal
structure). Again, the calculated magnetic moment of Cr has about
the same magnitude. A ground state with an anti-ferromagnetic or
ferrimagnetic order of the Cr atoms could not be verified in either
of both structures. Therefore, those structure can be excluded as
possible candidates for explanation of the reduction of the magnetic
moment. This is in accordance with the X-ray diffraction data where
no additional Bragg peaks were observed. Extra (001) Bragg peaks are
expected for both structures, for example. However, it is worthwhile
to note that the compound still exhibited HMF character in the
$cP16$ structure.

More promising are calculations for the $X$-structure
($F\:\overline{4}3m$) \cite{BP71}. (Note that the Pearson code of
this structure is $cF16$ like for $L2_1$.) This structure was also
considered among several phases with random anti-site disorder by
Feng \etal \cite{FRW01} investigating the physical properties of
Fe$_2$VAl. It is similar to the $C1_b$ Half-Heusler phase but with
the vacancies filled up in a different way compared to the $L2_1$
structure.  The $X$-structure consists of successive CoCr and CoAl
(001)-planes.

The calculation reveals for the $X$-structure an anti-ferromagnetic
ordering of the Cr atoms with respect to Co and thus a reduced
overall magnetic moment. The magnetic moments derived for the
different structures are summarized in table \ref{Tab3}.

\begin{table}
    \caption{Structural dependence of the magnetic moments in Co$_2$CrAl.\\
    Note: The sum of the magnetic moments of the two individual Co atoms
    and average values for Cr and Al are given to make different structures comparable.
    The total spin magnetic moment is given per formula unit.}
    \begin{center}
        \begin{tabular}{llcccc}
       &                     &        &       & $m\:[\mu_{B}]$ & \\
       & Structure           & Co$_2$ &  Cr   &  Al   & tot   \\
       \hline
$L2_1$ & $F\:m\overline{3}m$ & 1.42   &  1.63 & -0.05 & 2.98  \\
$X$    & $F\:\overline{4}3m$ & 1.84   & -0.83 & -0.02 & 0.91  \\
$cP16$ & $P\:m\overline{3}m$ & 1.69   &  1.48 & -0.05 & 3.00  \\
$tP4$  & $P\:4/mmm$          & 1.82   &  1.47 & -0.05 & 3.19  \\
       & Experiment          & 1.11   &  0.19 &   -   & 1.30
        \end{tabular}
    \end{center}
    \label{Tab3}
\end{table}

The calculated, element specific and total magnetic moments are
compared to experimental values in table \ref{Tab3}. For Co the sum
of both atoms is given. The average of the calculated values for Cr
and Al are given for the structures with inequivalent atomic
positions.

A more detailed analysis of the $X$-structure shows not only an
anti-ferromagnetic order of Cr but also an enhancement of the Co
magnetic moments. Their values are $0.94\mu_B$ in CoAl and
$0.76\mu_B$ in CoCr (001)-planes. Evidently the shortest distance
between two Co atoms is smaller in the $X$-structure compared to the
$L2_1$-structure (see table \ref{Tab4}). The shortest Co-Cr and
Cr-Cr distances stay the same in both structures.

The enhancement of the Co moments in all three non-$L2_1$ structures
does not allow to explain the experimentally observed moments
directly because the measured value is already smaller as the
calculated moment of the $L2_1$ structure. This is mainly a property
of the particular sample investigated here. Other samples exhibited
higher overall magnetic moments.

The gap in the minority band structure is closed in the
$X$-structure, that means it is not longer a half-metallic
ferromagnet. The vanishing of the gap was also found for Co$_2$FeAl
in the $X$-structure, but with the Fe atoms aligned
ferromagnetically with respect to Co. It is interesting to note that
the $cP16$ structure still exhibits the gap for both pure compounds.
From this observation it can be concluded that the existence of the
gap is directly related to the occurrence of Co$_2$ and mixed CrAl
(001)-planes and that the $L2_1$-structure is not the only
possibility for the presence of a HMF gap in Heusler-like compounds
X$_2$YZ.

\begin{sidewaystable}
     \caption{Site symmetries in Co$_2$CrAl. \\
     PM := paramagnetic order, FM := ferromagnetic order with $B\|[001]$.
     The nearest distance $d_{AB}$ between two atoms A and B is given in \AA.
     The second and $4^{th}$ lines for each structure give the Wyckoff positions
     of the atoms in the PM and FM state.}

    \small
    \begin{center}
        \begin{tabular}{lllcccccccccc}
    &                  &           & Co       & Cr       & Al       & $d_{CoCo}$ & $d_{CoCr}$ & $d_{CrCr}$ & $d_{CrAl}$
    & $\Gamma$         & $\Delta, (\Lambda$) [0,0,z] & $X, (Z)$ \\
\hline \\
$L2_1$ & PM  & $F\:m\overline{3}m$ & $T_d$    & $O_h$    & $O_h$    & 2.86 & 2.48 & 4.05 & 2.86
    & $O_h$            & $C_{4v}$ & $D_{4h}$ \\
$cF16$ & & & 8c & 4a & 4b &  & & & & & & \\

       & FM  & $I\:4/m$            & $S_4$    & $C_{4h}$ & $C_{4h}$ &&&&
    & $D_{4h}(C_{4h})$ & $C_{4v}(C_{4})$ & $D_{4h}(C_{4h})$ \\
    & &  & 4d & 2a & 2b &  & & & & & & \\
\hline \\
$X$    & PM  & $F\:\overline{4}3m$ & $T_d$    & $T_d$    & $T_d$    & 2.48 & 2.48 & 4.05 & 2.48
    & $T_d$            & $C_{2v}$ & $D_{2d}$\\
$cF16$ & & & 4a,4c & 4b & 4d &  & & & & & & \\

       & FM  & $I\:\overline{4}$   & $S_4$    & $S_4$    & $S_4$    &&&&
    & $D_{2d}(S_4)$    & $C_{2v}(C_{2})$ & $D_{2d}(S_4)$\\
    & &  & 2a,2d & 2b & 2c &  & & & & & & \\
\hline \\
$cP16$ & PM  & $P\:m\overline{3}m$ & $C_{3v}$ & $O_h, D_{4h}$    & $D_{4h}, O_h$    & 2.86 & 2.86 & 2.48 & 2.86
    & $O_h$ & $C_{4v}$ & $D_{4h}$ \\
    & & & 8g & 1a,3d & 3c,1b & & & & & & & \\

       & FM  & $P\:4/m$            & $C_1$    & $C_{4h}, C_{2h}, C_{2h}$ & $C_{2h}, C_{2h}, C_{4h}$ &&&&
    & $D_{4h}(C_{4h})$ & $C_{4v}(C_4)$ & $D_{4h}(C_{4h})$ \\
    & &  & 8l & 1a,2e,1b & 1c,2f,1d & & & & & & & \\
\hline \\
$tP4$  & PM  & $P\:4/mmm$          & $C_{4v}$ & $D_{4h}$ & $D_{4h}$ & 2.86 & 2.48 & 2.86 & 2.86
    & $D_{4h}$         & $C_{4v}$ & $D_{4h}$ \\
    & & & 2h & 1a & 1b & & & & & & & \\

       & FM  & $P\:4/m$            & $C_4$    & $C_{4h}$ & $C_{4h}$ &&&&
    & $D_{4h}(C_{4h})$ & $C_{4v}(C_4)$ & $D_{4h}(C_{4h})$ \\
    & & & 2h & 1a & 1b &  & & & & & & \\
        \end{tabular}
    \normalsize
    \end{center}
    \label{Tab4}
\end{sidewaystable}

This finding can be understood easily considering symmetry. The
magnetization will reduce the symmetry. Applying any special
orientation along one of the principal axes (e.g.: [001]) will
reduce the symmetry from $F\:m\overline{3}m$ to $I\:4/mmm$, at
least. This is in accordance with the fact that the $O_h$ point
group cannot describe ferromagnetic order. The properties of the
electron spin will cause a further reduction to $I\:4/m$, as
vertical mirror operations would change the sign of the spin. In
particular, the $\Gamma$-point of the ferromagnetic Heusler
compounds will belong to the $D_{4h}(C_{4h})$ colour group and not
longer to $O_h$ like in the paramagnetic state. The point group in
brackets assigns the {\it magnetic} symmetry with removed vertical
mirror planes. The $X$-point becomes $Z$ ($D_{4h}(C_{4h})$) and the
point group symmetry of the $\Lambda$-direction (formerly $\Delta$)
is reduced to $C_{4v}(C_4)$ in the ferromagnetic case.

The $\Gamma$ and $Z$ points of the $cP16$ or $tP4$ structures, as
representatives for Cr-Al disorder, have in the ferromagnetic state
{\it also} $D_{4h}(C_{4h})$ symmetry. Therefore, those structures
are expected to behave {\it similar} like the $L2_1$-structure.
However, the local symmetry of the atomic sites in the three
structures is different what may explain the vanishing of the HMF
gap in the $tP4$ structure.

The $\Gamma$ and $Z$ points of the $X$-structure, as representative
for Co-Cr anti-site disorder, have in the ferromagnetic state the
{\it lower} $D_{2d}(S_{4})$ point group symmetry. Therefore, that
structure is expected to behave {\it different} from the
$L2_1$-structure. Indeed, the most pronounced difference is the
local environment of the atoms in the (001)-planes.

The local site symmetries of the atoms are summarized in table
\ref{Tab4}. It displays the symmetries of atoms and high symmetric
points of the Brillouin-zone for the different structures for
paramagnetic (PM) and ferromagnetic (FM) order. The direction of
magnetization was chosen to be along the $z$-axis for FM order.  Cr
and Al atoms occupy in the $cP16$ structure two inequivalent sites
with different local symmetry. The distances between various atoms
are given for a fixed lattice parameter $a$. The $X$-point becomes
$Z$ and the $\Delta$-direction becomes $\Lambda$ in tetragonal
symmetry. The point group symmetries of the Brillouin-zone may serve
to avoid confusion about the irreducible representation of bands
being different in the PM and FM state.

The Cr magnetic moment is also too small in the mixed
Co$_2$Cr$_{1-x}$Fe$_{x}$Al alloy, as seen from figure \ref{fig7}.
The Fe magnetic moment comes close to the value expected from the
calculation. This observation points not only on a site-disorder but
also on the possibility of phase separation resulting in Fe and Cr
rich grains of the polycrystalline sample. The overall magnetic
moment comes close to the calculated value as result of the much
higher value for Fe compared to Cr. From this observation it is
clear that measuring only the total moment is not enough to
characterize these alloys completely.

Similar calculations were performed for the Half-Heusler compound
NiMnSb in order to explain a missing full spin polarization in that
compound. In the XYZ $C1_b$-structure one has a series of pure X
followed by mixed YZ (001)-planes. The calculations were performed
for NiMnSb, MnSbNi, and SbNiMn, keeping the lattice parameter fixed.
The major difference between the three types is the local
environment of the Ni atoms in the (001) planes. Only in the first
type, the alternating (001)-planes contain either purely Ni or mixed
(MnSb) layers. The exchange of Ni and Mn or Sb resulted in the loss
of the HMF character and in turn in a reduction of the spin
polarization at the Fermi-energy. Indeed, such an intermixing must
not prevail through the hole crystal. It is just enough to have
disorder in regions close to the surface or interface in order to
explain a reduced spin polarization in spectroscopic methods. These
findings are in agreement with the results of Orgassa \etal
\cite{OFS99} for random disorder in NiMnSb.

The assumption of doped, ordered compounds may not hold in every
case, especially if considering non half- or non quarter-integer
fractions for the iron concentration $x$. Calculations concerning
the spectroscopic properties for more general, non-rational iron
concentration, resulting in random alloys, are in preparation.
However, the results found here for superstructures are in well
agreement to those of Miura \etal \cite{MNS04} for random alloys.

\section{Summary}

The electronic structure of the pure and doped Heusler compound
Co$_2$Cr$_{1-x}$Fe$_{x}$Al with varying iron content ($x$) was
calculated by means of different theoretical methods. Element
specific magnetic moments were determined from MCD measurements at
the Cr, Fe, and Co $L_{2,3}$ absorption edges of the Heusler alloys
Co$_2$CrAl, Co$_2$Cr$_{0.6}$Fe$_{0.4}$Al, and Co$_2$FeAl.

The calculations revealed a ferromagnetic coupling between the $3d$
atoms as well as an anti-ferromagnetic alignment of the Al magnetic
moments with respect to the moments of the $3d$ elements. However,
the Al moments are very small and induced by the surrounding
polarized atoms. The calculations predict the
Co$_2$Cr$_{1-x}$Fe$_{x}$Al compound to be a half-metallic
ferromagnet. The size of the minority gap ranges from $100meV$ to
$800meV$. The smallest band-gap around the Fermi-edge in the
minority bands was found for the compound with $x=1$. Co$_2$CrAl
and, more pronounced, Co$_2$FeAl turned out to have an indirect
$\Gamma - X$ gap. The mixed compounds exhibit a direct gap at the
$\Gamma$ point being caused by the reduced symmetry.

It was shown that the origin of the minority gap in Co$_2$CrAl is
the geometrical structure and local symmetry of the atoms. It
appears in the $L2_1$-structure with successive CrAl and Co$_2$
(001)-planes but not in the $X$-structure with successive CoCr and
CoAl (001)-planes.

In summary, it was shown how theoretical methods can be used to
design new materials with predictable magnetic properties.

\section{Acknowledgments}

The authors thank all members of NSRRC (Hsinchu, Taiwan) for their
help during the beamtimes. G.H.F. and S.W. are very grateful to
Yeukuang Hwu (Academia Sinica, Taipei) and his group for support
during the experiments in Taiwan.

Financial support by the DFG (FG 559), DAAD (03/314973 and
03/23562), and PROBRAL (167/04) is gratefully acknowledged.

\newpage
\bibliographystyle{unsrt}
\bibliography{Fecher_DesignOfMagMat}

\end{document}